\documentclass[20pt]{article}
\usepackage{spconf,amsmath,graphicx}
\usepackage{longtable}
\usepackage{amssymb, booktabs}
\usepackage{anyfontsize}
\usepackage{multirow}
\usepackage{caption}
\usepackage[table]{xcolor}
\usepackage{float}
\usepackage{url}
\usepackage{cite}
\DeclareCaptionFont{mycaptionsize}{\fontsize{9pt}{10pt}\selectfont}
\title{DISPATCH: Distilling Selective Patches for Speech Enhancement}
\name{Dohwan Kim$^1$ and Jung-Woo Choi$^1$\sthanks{Corresponding author.}}
\address{
School of Electrical Engineering, KAIST, Daejeon, Republic of Korea \\
\{dohwan, jwoo\}@kaist.ac.kr
}

\begin{document}

\ninept
\maketitle
\begin{abstract}
In speech enhancement, knowledge distillation (KD) compresses models by transferring a high-capacity teacher’s knowledge to a compact student. However, conventional KD methods train the student to mimic the teacher’s output entirely, which forces the student to imitate the regions where the teacher performs poorly and to apply distillation to the regions where the student already performs well, which yields only marginal gains. We propose Distilling Selective Patches (DISPatch), a KD framework for speech enhancement that applies the distillation loss to spectrogram patches where the teacher outperforms the student, as determined by a Knowledge Gap Score. This approach guides optimization toward areas with the most significant potential for student improvement while minimizing the influence of regions where the teacher may provide unreliable instruction. Furthermore, we introduce Multi-Scale Selective Patches (MSSP), a frequency-dependent method that uses different patch sizes across low- and high-frequency bands to account for spectral heterogeneity. We incorporate DISPatch into conventional KD methods and observe consistent gains in compact students. Moreover, integrating DISPatch and MSSP into a state-of-the-art frequency-dependent KD method considerably improves performance across all metrics. \footnote{\url{https://github.com/rlaehghks5/DISPATCH}}

\end{abstract}
\vspace{-1.5pt}
\section{Introduction}
\vspace{-1.5pt}
\label{sec:intro}
While deep neural networks have demonstrated strong performance in speech enhancement \cite{defossez2020real, hu2020dccrndeepcomplexconvolution, lee2023deft}, their high computational demands pose significant challenges for practical applications, particularly in resource-constrained settings like on-device AI, where hardware limitations, power consumption, and monetary costs are critical concerns. Consequently, model compression has become a crucial research direction for developing computationally efficient speech enhancement models. Several compression strategies have been explored, including pruning \cite{han2015learningweightsconnectionsefficient}, which removes redundant weights; quantization \cite{courbariaux2014training}, which reduces the numerical precision of model parameters; and Knowledge Distillation (KD) \cite{hinton2015distilling, park2019relational, mansourian2025comprehensive}. 
This paper focuses on KD, a teacher-student paradigm in which a compact student model is trained to mimic a larger teacher model, thereby inheriting the teacher's knowledge.

KD has been actively explored in the speech domain, such as speaker verification \cite{mingote2023class, Truong_2024}, speech separation \cite{8489456, zhang2021teacherstudentmixitunsupervisedsemisupervised}, and speech enhancement \cite{subband-kd, 10502149, abc-kd, nathoo2024two, dfkd}. A prominent line of work is response-based KD, which trains a student to mimic the teacher's output \cite{mingote2023class, 8489456, zhang2021teacherstudentmixitunsupervisedsemisupervised, subband-kd, dfkd}. This output can be diverse, ranging from logits or posterior probabilities to time-frequency output. Some approaches \cite{Truong_2024, 10502149, abc-kd} combine response-based and feature-based KD to align intermediate representations, leveraging complementary information.

In speech enhancement, distilling time–frequency output has become a prominent response-based KD strategy because this approach can effectively capture the rich time-frequency structure of speech. This approach has given rise to notable frequency-dependent methods, such as Subband-KD \cite{subband-kd} and Dynamic Frequency-Adaptive Knowledge Distillation (DFKD) \cite{dfkd}. DFKD addresses the limitation of Subband-KD arising from static band partitioning by introducing a frequency adapter that dynamically partitions the spectrogram into low- and high-frequency regions and optimizes each region with distinct losses. This dynamic frequency-dependent approach allows DFKD to significantly outperform Subband-KD and ABC-KD \cite{abc-kd}, a hybrid of response- and feature-based distillation. Although DFKD has shown strong performance, it remains limited by relying on the entire teacher output, which may contain low‑quality time–frequency regions.

KD generally assumes that the teacher provides reliable supervision. However, this premise overlooks two critical issues in practice. The first limitation concerns the student's existing proficiency. Distilling knowledge in regions where the student has already converged yields minimal benefit. The second challenge stems from the inherent imperfections of the teacher model, which may result from overfitting to the training dataset. Therefore, forcing the student to learn from flawed regions in the teacher’s output is detrimental, as it leads the student to replicate the teacher's weakness instead of its strengths. As a result, indiscriminate distillation not only diverts the student's focus from the regions where the teacher's guidance is most needed but also, more importantly, can cause the student to imitate the teacher's erroneous output, leading to a degradation in performance. Therefore, it is crucial to adopt a selective strategy that allows the student to learn only from beneficial supervision. Although several adaptive strategies have been proposed, such as importance-weighted KD \cite{kim2024maximizing, ham2024difficulty} and uncertainty-aware KD \cite{jin2020uncertaintyawaremultishotknowledgedistillation, huang2023uncertainty}, they still presuppose that the teacher is uniformly superior to the student for a given input. Consequently, they inevitably transfer local artifacts even under reduced weights, neglecting that the student may locally outperform the teacher.


\begin{figure*}
    \centering
    \includegraphics[width=1.0\textwidth]{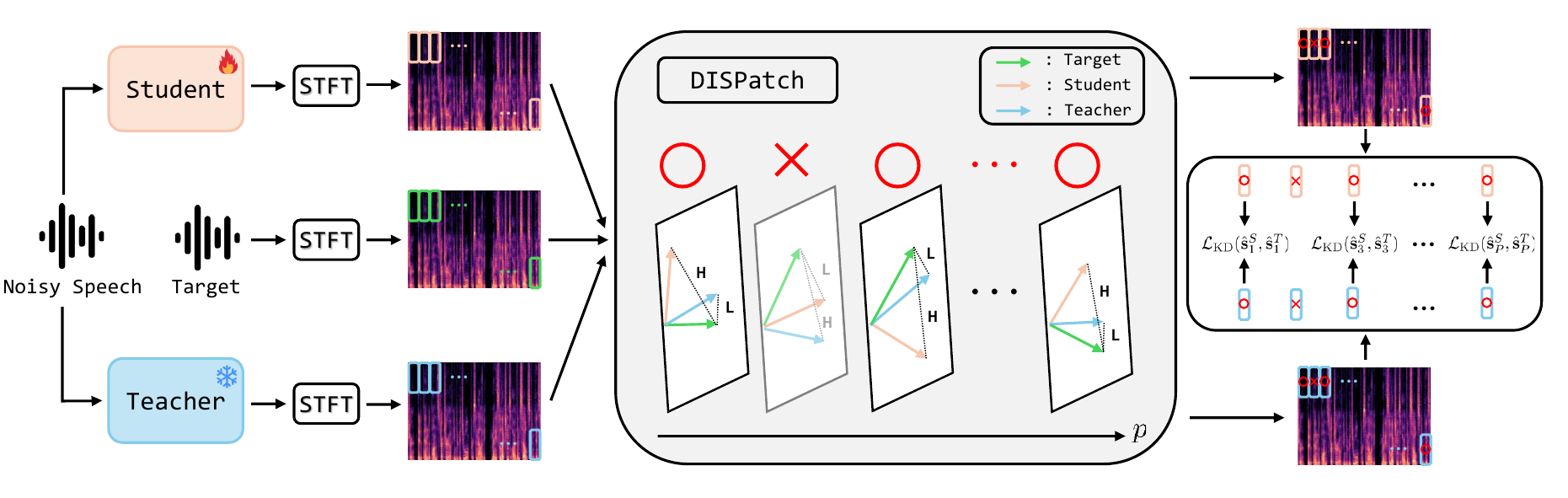}
    \caption{\textbf{Overview of DISPatch framework.} DISPatch framework separates spectrograms into patches and selectively applies KD loss to those exhibiting a high Knowledge Gap, where the teacher's error is low (L) and the student's is high (H). Patch size: $2C\times N \times1$ .}
    
    \label{fig:1}
\end{figure*}
Recent work on large language models has demonstrated that their accelerated convergence and improved performance can be achieved through selective training strategies \cite{lin2024rho}. This approach involves training the model exclusively on useful tokens, identified by a reference model trained on a high-quality dataset. Inspired by this principle, we propose Distilling Selective Patches (DISPatch). This strategy distills only the most useful spectrogram patches, leveraging KD's selectivity and mitigating the limitations discussed above. The overall architecture of DISPatch is illustrated in Fig.\ref{fig:1}. We introduce a Knowledge Gap Score (KGS) to enable a more targeted transfer of the teacher’s knowledge by identifying and distilling only the most valuable patches, rather than imposing the KD loss across the entire spectrogram. DISPatch explicitly ignores regions where the teacher is unreliable or where the student is already proficient. Moreover, recognizing that speech characteristics vary across frequencies, we introduce Multi-Scale Selective Patches (MSSP). This is a frequency-dependent strategy that adapts the distinct patch sizes to different frequency regions for more tailored distillation. We show consistent gains by applying DISPatch to response-based KD, and further improvements when integrating DISPatch and MSSP into DFKD.

\section{METHODOLOGY}

\subsection{Not All Spectrogram Regions Are Equal} 
\label{sec:2.1}




To demonstrate the presence of informative regions in the output of teacher and student models, we follow the approach proposed in \cite{lin2024rho} and analyze spectrogram patches on the DNS2020 Challenge dataset using ConvTasNet \cite{kadioglu2020empiricalstudyconvtasnet}. For this preliminary analysis, we set up a teacher and a student model; their detailed configurations are described in Section \ref{sec:3.2}. Let \( \mathbf{S}, \hat{\mathbf{S}} \in \mathbb{R}^{2C \times F \times T} \) denote the clean target and the estimated complex spectrogram, respectively. Here, \(C\), \(F\), and \(T\) denote the numbers of channels, frequency bins, and time frames. To quantify the error on the local level, we separate the complex spectrogram into $P$ patches: $\mathbf{s}_p,~\hat{\mathbf{s}}_p \in \mathbb{R}^{2C \times N \times 1}$ ($p\in \{ 0,\, \cdots,\, P-1 \}$), where $N=FT/P$ is the number of frequency bins in each patch and each patch has a single time frame. In practice, we use zero padding along the frequency axis so that $F$ becomes a multiple of $N$. Specifically, padding is used only at the high-frequency end of the spectrogram, and padded bins are ignored in all computations.
Then, we define the patch-wise magnitude error:
\begin{equation}
e_{p} = \left\| |\mathbf{s}_{p}|  -  |\hat{\mathbf{s}}_{p}| \right\|_2^2
\end{equation}
where $|\mathbf{s}_p|$ and $|\hat{\mathbf{s}}_{p}|$ are the magnitudes of the $p$-th patch in the target and estimated complex spectrograms, respectively. Then, for each patch of a given sample in the training data, we record the error every 1k training steps to construct an error sequence $(e_{p}^{(0)},\,e_{p}^{(1)},\,\cdots,\,e_{p}^{(J)})$ for $J+1$ observation points. To assess error convergence during training, we perform a linear regression by fitting a first-order polynomial as shown below:

\begin{figure}
\begin{minipage}[b]{.48\linewidth}
  \centering
  \centerline{\includegraphics[width=4.2cm]{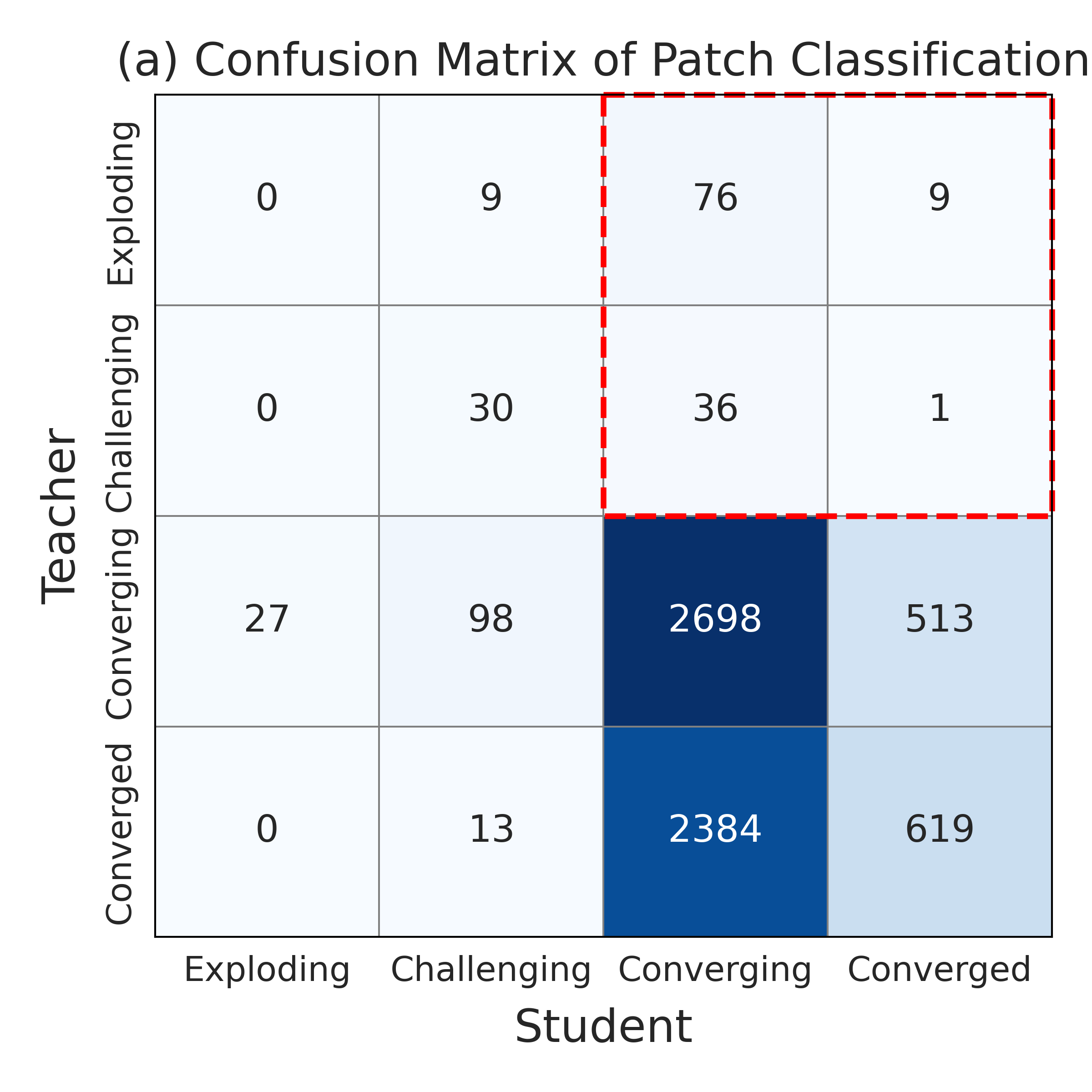}}
\end{minipage}
\begin{minipage}[b]{.48\linewidth}
  \centering
  \centerline{\includegraphics[width=4.2cm]{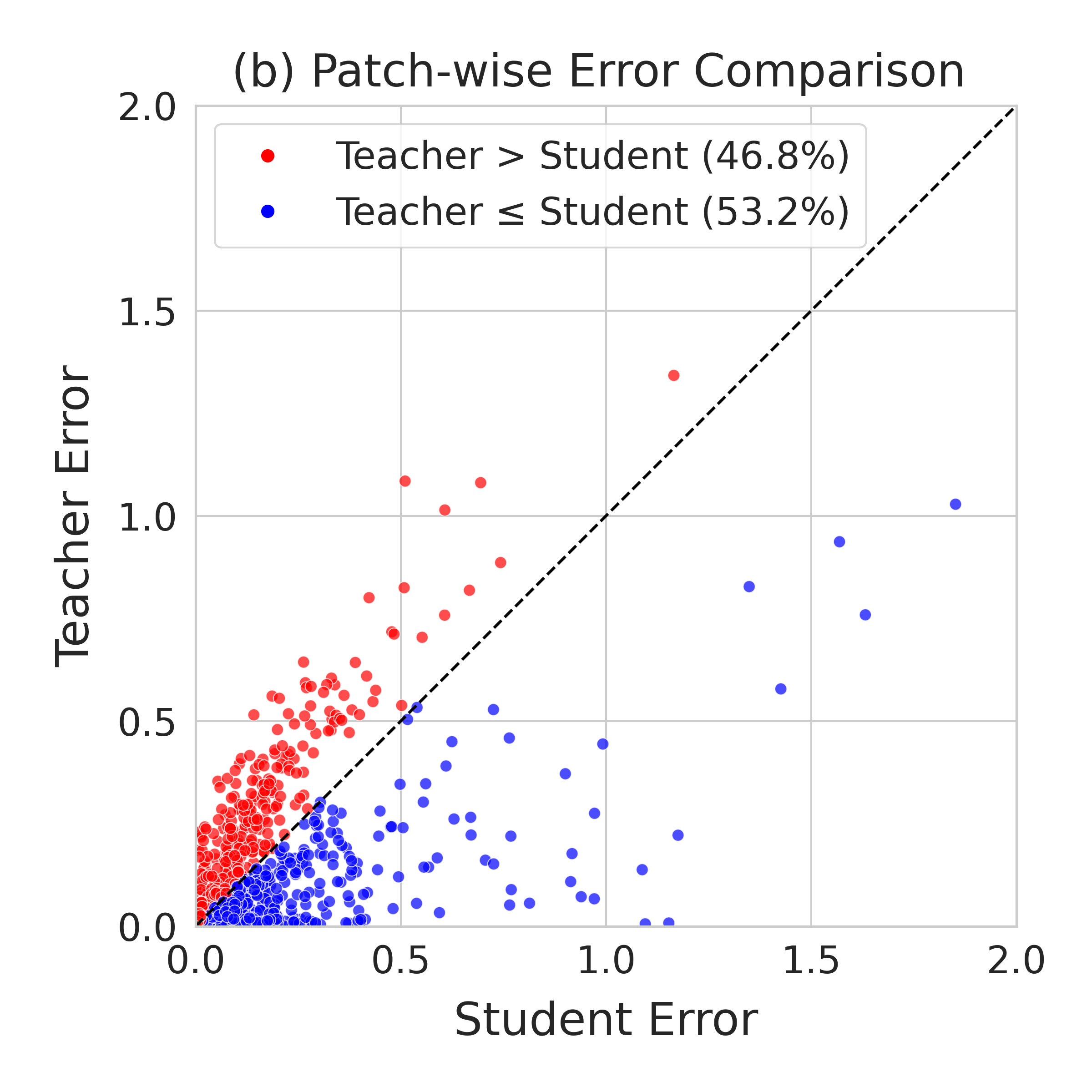}}
\end{minipage}
\caption{\textbf{Analysis of patch-level teacher-student misalignment.}  (a) Confusion matrix of patch classification. (b) Scatter plot comparing patch-wise errors between the teacher and the student.}
\label{fig:2}
%
\vspace{-13pt}
\end{figure}

\begin{equation}
f(a_p,b_p) =  \min_{a_p,b_p} \sum_{j=0}^{J} \left| e_{p}^{(j)} - (a_p j + b_p) \right|^2
\end{equation}
For each patch, the change between the first and last points in the fitted line $\Delta e_p = a_p\cdot J$ is calculated to determine the convergence of the error. 
Additionally, we compute the average error across 
all patches at the final iteration point, given by:
\begin{equation}
    e_{\mathrm{mean}} = \frac{1}{P} \sum_{p=0}^{P-1} e_p^{(J)}
\end{equation}
Based on $\Delta e_p$ and  $e_{\mathrm{mean}}$, we classify each patch into one of four categories. 
Patches are labeled \textit{Converging} if $\Delta e_p < -0.01$, and \textit{Exploding} if $\Delta e_p > 0.01$. The patches with stable error $(-0.01 \le \Delta e_p \le 0.01)$ are labeled \textit{Converged} if $e^{(J)}_{p} \le e_{\mathrm{mean}}$, and \textit{Challenging} if $e^{(J)}_{p} > e_{\mathrm{mean}}$. Note that decision thresholds other than 0.01 (e.g., 0.02 or 0.03) result in similar patch distributions.

Contrary to the common assumption that the teacher consistently provides superior supervision, our patch-level analysis of a single speech sample reveals misalignment between the teacher and student. Fig.\ref{fig:2} (a) visualizes the patch classification mismatch between the two models. Although the student has stabilized in many regions (\textit{Converged} or \textit{Converging}), a significant subset exists where the teacher remains unstable (\textit{Exploding} or \textit{Challenging}) despite the student’s proficiency (highlighted in the red box). Furthermore, Fig.\ref{fig:2} (b) confirms that at the $J$-th point, the teacher’s error exceeds the student’s in 46.8\% of patches. This empirical evidence exposes two critical risks of indiscriminate distillation: the transfer of harmful artifacts by forcing the student to mimic the teacher’s flawed regions, and the diversion of the student’s focus to regions where the student already excels, rather than where distillation is truly needed.



\label{sec:Mtehod}
\subsection{DISPatch:  Distilling Selective Patches} 
Based on our previous analysis, we propose Distilling Selective Patches (DISPatch), which selectively distills knowledge from patches where the teacher outperforms the student. To achieve this, we first define a patch-wise error metric $\varepsilon(\mathbf{s}_p, \, \hat{\mathbf{s}}_p)$.
The patch-wise error metric $\varepsilon$ is designed to be flexible. It can be defined as a simple magnitude spectrogram distance (e.g., L1 or L2) or as a more advanced objective, such as the one proposed in DFKD \cite{dfkd}. Using this metric, we compute the patch-wise error for both the student and teacher models with respect to the target speech:
\begin{equation}
    E_{p}^{S} = \varepsilon(\mathbf{s}_p, \, \hat{\mathbf{s}}_p^S), ~~ E_{p}^{T} = \varepsilon(\mathbf{s}_p, \, \hat{\mathbf{s}}_p^T)
\end{equation}
where superscripts S and T denote the student and the teacher, respectively. To quantify the knowledge gap between the student and the teacher in each patch, we introduce the Knowledge Gap Score (KGS), an intuitive metric designed to focus distillation on patches where the teacher's guidance is most beneficial:
\begin{equation}
KGS_{p} = E^{S}_{p}-E^{T}_{p}.
\end{equation}
A higher $KGS_p$ indicates a greater knowledge disparity, suggesting that the teacher's knowledge in that patch is particularly valuable for distillation. Consequently, we selectively transfer knowledge only from patches with high $KGS_p$values. The distillation loss is applied exclusively to the top $k$\% of patches ranked by their $KGS_p$:
\begin{equation}
\mathcal{L}_{\mathrm{DISPatch}} = \frac{1}{P \cdot k\%} \sum_{i=0}^{P-1} I_{k\%}({KGS}_p) \cdot \mathcal{L}_{\mathrm{KD}}(\hat{\mathbf{s}}_p^S, \hat{\mathbf{s}}_p^T)
\end{equation}
where $\mathcal{L}_{\mathrm{KD}}$ is a KD loss for $p$-th patch, and $I_{k\%}$ is the indicator function defined as:
\begin{equation}
I_{k\%}({KGS}_p) =
\begin{cases}
1 & \text{if } {KGS}_p \text{ ranks in the top } k\%, \\
0 & \text{otherwise}
\end{cases}
\end{equation}
DISPatch allows for the dynamic selection of critical patches within each speech sample, imposing the KD loss to match the teacher's output only on a selected informative subset. In summary, selecting patches based on high $KGS_p$ enables the model to avoid suboptimal distillation by naturally excluding regions where the teacher is unreliable or the student already performs well.

\subsection{MSSP: Multi-Scale Selective Patches} 

A single, uniform patch size is suboptimal for DISPatch because it fails to account for the heterogeneous characteristics of speech signals across frequency bands. Acoustic phonetics shows that low-frequency regions are typically dominated by fine-grained structured information such as formants, which are crucial for vowel identity \cite{diehl2008acoustic}. In contrast, high-frequency regions often contain broader, less-structured energy patterns, such as those found in fricatives \cite{jongman2000acoustic}. Therefore, we propose Multi-Scale Selective Patches (MSSP), which uses a smaller patch size for low frequencies and a larger one for high frequencies to account for these heterogeneous characteristics.

To implement MSSP, we identify the dynamic crossover point by adopting the frequency adapter from DFKD \cite{dfkd}, which determines the boundary for each time frame. DFKD dynamically partitions output into low- and high-frequency bands via the frequency adapter and uses frequency-dependent KD objectives to balance speech preservation at low frequencies and noise suppression at high frequencies. In practice, a phase-only cosine loss is used in the low-frequency band, while the high-frequency band uses a combination of amplitude L2 and phase losses. The combination of MSSP and DFKD is conceptually aligned, as both adopt frequency‑dependent objectives. Accordingly, we integrate MSSP into DFKD to evaluate the effect of employing distinct patch sizes across low and high frequencies.

\vspace{-2pt}
\subsection{Training Setup with DISPatch}
 The overall training objective, $\mathcal{L}_{\mathrm{total}}$, is defined as a weighted sum of speech enhancement loss ($\mathcal{L}_{\mathrm{SE}}$) and our proposed selective distillation loss ($\mathcal{L}_{\mathrm{DISPatch}}$) as follows:
\begin{equation}
\mathcal{L}_{\mathrm{total}} = \alpha \cdot \mathcal{L}_{\mathrm{SE}} + (1-\alpha) \cdot \mathcal{L}_{\mathrm{DISPatch}}
\end{equation}
where $\alpha$ is a hyperparameter that balances the contribution of the two loss terms. In our experiments, we set $\alpha$ to 0.5, giving equal weight to direct supervision and selective knowledge distillation. To avoid overfitting the weighting to any particular method, we fix $\alpha$ across all models and different KD objectives, rather than tuning it per method. For $\mathcal{L}_{\mathrm{SE}}$, we adopt the objective function originally proposed for each respective model.


\begin{table*}[ht]
\vspace{-6pt}
\fontsize{7.2pt}{8.15pt}\selectfont
\centering
\caption{Comparison of DISPatch-Augmented Knowledge Distillation Methods on DCCRN-CL and ConvTasNet}
\label{tab:my_table1}
\begin{tabular}{c|c|c|c@{\hspace{6pt}}c@{\hspace{7pt}}@{\hspace{3pt}}c@{\hspace{3pt}}c@{\hspace{5pt}}c@{\hspace{5pt}}@{\hspace{5pt}}c@{\hspace{3pt}}c@{\hspace{5pt}}c@{\hspace{5pt}}}
\toprule [1pt]
\multirow{2}{*}[-1pt]{\textbf{Model}} & \multirow{2}{*}[-1pt]{\textbf{Param.}} & \multirow{2}{*}{\textbf{MAC/s}} & \multirow{2}{*}[-1pt]{\textbf{Methods}} & \multirow{2}{*}[-1pt]{\textbf{DISPatch}} & \multicolumn{3}{c}{\textbf{DNS2020-test}} & \multicolumn{3}{c}{\textbf{VoiceBank+DEMAND}} \\
& & & & & \textbf{WB-PESQ} & \textbf{NB-PESQ} & \textbf{STOI(\%)} & \textbf{WB-PESQ} & \textbf{NB-PESQ} & \textbf{STOI(\%)} \\ 
\midrule

\multirow{1}{*}{ConvTasNet (Teacher)} & 9.71M & 8.05G & Scratch  & X & 2.792 & 3.337 & 96.71 & 2.770 & 3.582 & 94.12 \\
\midrule
\rowcolor{gray!30} \multirow{8}{*}{ConvTasNet (Student)}& \multirow{8}{*}{1.75M} & \multirow{8}{*}{0.57G} & Scratch & X & 2.563 & 3.122 & 96.05 & 2.585 & 3.424 & 93.44 \\
\cline{4-11}
& & & L1 & X & 2.584 & 3.135 & 96.11 & 2.584 & 3.429 & 93.57 \\ 
& & & L1 & O & \textbf{2.602} & \textbf{3.165} & \textbf{96.16} & \textbf{2.614} & \textbf{3.450} & \textbf{93.79} \\
\cline{4-11}
& & & L2 & X & 2.563 & 3.127 & 95.98 & 2.556 & 3.400 & 93.37 \\ 
& & & L2 & O & \textbf{2.602} & \textbf{3.157} & \textbf{96.10} & \textbf{2.603} & \textbf{3.431} & \textbf{93.39} \\
 \cline{4-11}
& & & DFKD & X & 2.623 & 3.170 & 96.08 & 2.583 & 3.458 & 93.70 \\ 
& & & DFKD & O & 2.650 & 3.208 & 96.12 & 2.589 & 3.490 & 93.81 \\ 
\rowcolor{red!15} & & & DFKD + MSSP(10/40)& O & \textbf{2.677} & \textbf{3.226} & \textbf{96.17} & \textbf{2.618} & \textbf{3.505} & \textbf{94.02} \\ 
\midrule  [1pt]
\multirow{1}{*}{DCCRN-CL (Teacher)} & 9.22M & 15.33G & Scratch & X & 2.764 & 3.318 & 96.38 & 2.655 & 3.591 & 92.94 \\ 
\midrule
\rowcolor{gray!30} \multirow{8}{*}{DCCRN-CL (Student)} & \multirow{8}{*}{2.24M} & \multirow{8}{*}{3.84G} & Scratch & X & 2.691 & 3.247 & 96.08 & 2.582 & 3.482 & 92.96 \\
\cline{4-11}
 & & & L1         & X & 2.660  & 3.236 & 96.07 & 2.631 & 3.454 & 92.98  \\ 
 & & & L1 & O & \textbf{2.709}  & \textbf{3.270}  & \textbf{96.13} & \textbf{2.678} & \textbf{3.497} & \textbf{93.20}  \\
\cline{4-11}
 & & & L2         & X & 2.690  & 3.241 & 95.96  & 2.587 & 3.446 & 92.96 \\
 & & & L2         & O & \textbf{2.726} & \textbf{3.283} & \textbf{96.08} & \textbf{2.675} & \textbf{3.486} & \textbf{93.04} \\
\cline{4-11}
 & & & DFKD       & X & 2.719 & 3.279 & 96.06  & 2.635 & 3.476 & 93.04  \\
 & & & DFKD       & O & 2.757 & 3.291 & \textbf{96.10} & 2.618 & 3.490  & 93.13 \\
\rowcolor{red!15} & & & DFKD + MSSP(10/40) & O & \textbf{2.758} & \textbf{3.301} & \textbf{96.10} & \textbf{2.719} & \textbf{3.535} & \textbf{93.29} \\ 
\bottomrule [1pt]
\end{tabular}
\end{table*}

\section{EXPERIMENTS}

\subsection{Datasets}

For training, we use the dataset from the DNS2020 challenge \cite{interspeech2020deepnoise}, which consists of more than 500 and 100 hours of clean speech and noise, respectively. The speech and noise are mixed at various Signal-to-Noise Ratios (SNRs) uniformly sampled from 0 to 20 dB, and all audio clips are 4-second segments sampled at 16 kHz. To demonstrate the effectiveness of DISPatch, we conduct evaluations on both the DNS2020 challenge test set and the VoiceBank+DEMAND \cite{botinhao2016investigating} test set, also sampled at 16kHz. 

\subsection{Implementation Details}
\label{sec:3.2}
We evaluate DISPatch using ConvTasNet (Deep w/ PReLU) \cite{kadioglu2020empiricalstudyconvtasnet} and DCCRN-CL \cite{hu2020dccrndeepcomplexconvolution}. For the implementation of ConvTasNet (Deep w/ PReLU), the teacher (9.71M) and student (1.75M) are configured with hyperparameters \{N, L, B, H, Sc, P, X, R\} set to \{512, 16, 128, 512, 128, 3, 8, 3\} and \{128, 40, 128, 256, 128, 3, 7, 2\}, respectively. The DCCRN-CL teacher (9.22M) and student (2.24M) are configured with encoder channels of \{64, 128, 256, 256, 512, 512\} and \{32, 64, 128, 128, 256, 256\}, respectively. Additionally, the teacher is configured with two complex LSTM layers and 256 hidden units, while the student has one layer and 128 units. Both models are trained for 100 epochs with an initial learning rate of $1 \times 10^{-3}$, which is halved when the validation loss does not improve for three consecutive epochs; training is stopped if no improvement is seen for ten epochs. We use Adam and apply gradient clipping with a maximum L2 norm of 5 and a batch size of 16.

For the analysis in Section \ref{sec:2.1}, we set the patch size $N$ to 20. We used the same setting for DISPatch with a single patch size and set the patch selection ratio $k$ to 80. We evaluated the effectiveness of DISPatch on response-based KD methods, including L1, L2, and DFKD. DISPatch is applied using the STFT with a Hann window of 32 ms and a hop size of 8 ms, corresponding to 75\% overlap. The additional computational cost of DISPatch is approximately 1.2\% per epoch when used with DFKD on ConvTasNet. All hyperparameters related to DISPatch are kept consistent across both models and all KD objectives. 



\vspace{-4pt}
\subsection{Experimental Results}
\vspace{-2pt}
\noindent Table \ref{tab:my_table1} indicates that magnitude distance-based KD methods (L1, L2) offer marginal gains over the scratch baseline and often result in performance degradation. In contrast, applying DISPatch consistently improves all metrics for both architectures, substantially outperforming the methods without DISPatch and surpassing the scratch baseline in most cases. The gains persist with DFKD, highlighting the broad effectiveness of focusing distillation on informative patches. These results clearly highlight the benefit of distilling from only the most informative patches. The most significant gains are achieved when integrating DFKD with both DISPatch and MSSP (10/40 for low/high), achieving the highest performance across both test sets. These results demonstrate that MSSP further amplifies the strengths of DFKD, which is a frequency-dependent method, by using asymmetric patch scales for low- and high-frequency bands. 

\begin{table}[t]
\centering
\fontsize{6.35pt}{7.5pt}\selectfont
\vspace{-8pt}
\caption{Performance Comparison of MSSP}
\label{tab:my_table2}
\begin{tabular}{@{\hspace{1pt}}c@{\hspace{2pt}}|@{\hspace{2pt}}c@{\hspace{2pt}}|@{\hspace{3pt}}c@{\hspace{3pt}}c@{\hspace{3pt}}c@{\hspace{3pt}}c@{\hspace{3pt}}c@{\hspace{3pt}}c@{\hspace{1pt}}}
\toprule [1pt]
\multirow{2}{*}[-1pt]{\textbf{Model}} & \multirow{2}{*}[-1pt]{\textbf{low/high}} & \multicolumn{3}{c}{\textbf{DNS2020-test}} & \multicolumn{3}{c}{\textbf{VoiceBank+DEMAND}} \\ 
& & \textbf{WB-PESQ} & \textbf{NB-PESQ} & \textbf{STOI(\%)} & \textbf{WB-PESQ} & \textbf{NB-PESQ} &\textbf{STOI(\%)} \\ \midrule 
\multirow{4}{*}{ConvTasNet}& 20/20 & 2.650 & 3.208 & 96.12 & 2.589 & 3.490 & 93.81 \\ 
& 10/20 & 2.630 & 3.191 & 96.01 & 2.555 & 3.470 & 93.66 \\ 
& 20/40 & 2.662 & 3.211 & 96.12 & \textbf{2.626} & 3.483 & 93.76 \\ 
& 10/40 & \textbf{2.677} & \textbf{3.226} & \textbf{96.17} & 2.618 & \textbf{3.505} & \textbf{94.02}\\
\midrule  [1pt]

\multirow{4}{*}{DCCRN-CL}& 20/20 & 2.757 & 3.291 & 96.10 & 2.618 & 3.490 & 93.13 \\ 
& 10/20 & 2.746 & 3.295 & 96.09 & 2.690 & 3.506 & 93.28 \\ 
& 20/40 & 2.745 & 3.298 & \textbf{96.11} & 2.668 & 3.506 & 93.16 \\ 
& 10/40 & \textbf{2.758} & \textbf{3.301} & 96.10 & \textbf{2.719} & \textbf{3.535} & \textbf{93.29} \\ 
\bottomrule [1pt]
\end{tabular}
\vspace{-12pt}
\end{table}

To validate the effectiveness of MSSP and identify a proper patch configuration, we compare DFKD with DISPatch using a uniform patch size (20/20) against DFKD incorporating MSSP and DISPatch, as detailed in Table \ref{tab:my_table2}. The values of patch sizes are chosen to test the impact of local (10) and broader (40) scopes for low and high frequencies against the uniform baseline (20). For both models and across two datasets, the configuration applying a smaller patch size in low frequencies and a larger patch size in high frequencies (10/40) achieves the highest scores, outperforming using only a small low-frequency patch (10/20) or only a large high-frequency patch (20/40) when compared to the uniform patch size (20/20). These results support our hypothesis that tailoring patch sizes to the distinct acoustic characteristics of different frequency bands enables a better enhancement of the student's capabilities.


For an ablation study on the components of KGS, we adopt objective L2 to align with the analysis setup in Section \ref{sec:2.1}, and we evaluate patch selection criteria where Top-$k$ and Bottom-$k$ denote selecting the highest and lowest $k$\% patches, respectively.
Each patch selection criterion is selected using one of the individual components of $KGS_p$($E^{S}_{p}$, $E^{T}_{p}$). 
Table \ref{tab: ablation KGS} shows that the proposed $KGS_p$ is the most effective selection criterion, whereas relying on a single component degrades performance.
First, the patches selected by Top-$k$ by $E^{T}_{p}$ or Bottom-$k$ by $E^{S}_{p}$ are not priorities for distillation, as they select patches with high teacher error and patches where the student’s error is already low. Second, although selecting patches with Bottom-$k$ by $E^{T}_{p}$ and Top-$k$ by $E^{S}_{p}$ are individually informative, $KGS_p$ still outperforms them. This gap occurs because each single component criterion fails to exclude its uninformative or harmful complement, respectively. For example, Bottom-$k$ by $E^{T}_{p}$ fails to filter out patches where the student is already proficient (low $E^{S}_{p}$), and Top-$k$ by $E^{S}_{p}$ fails to avoid patches where the teacher also performs poorly (high $E^{T}_{p}$). In contrast, $KGS_p$ selects patches based on the difference ($E^{S}_{p}-E^{T}_{p}$), inherently considering both counterparts to identify regions with the most significant potential for student improvement.

\label{sec:pagestyle}

\begin{table}[t]
\centering
\fontsize{6.35pt}{7.5pt}\selectfont
\vspace{-8pt}
\caption{Ablation Study on the Components of KGS}
\label{tab: ablation KGS}
\begin{tabular}
{@{\hspace{2pt}}c@{\hspace{2pt}}|c@{\hspace{2pt}}c@{\hspace{2pt}}c@{\hspace{4pt}}c@{\hspace{2pt}}c@{\hspace{2pt}}c@{\hspace{2pt}}}

\toprule [1pt]
\multirow{2}{*}[-1pt]{\textbf{Selection Criterion}} & \multicolumn{3}{c}{\textbf{DNS2020-test}} & \multicolumn{3}{c}{\textbf{VoiceBank+DEMAND}} \\
& \textbf{WB-PESQ} & \textbf{NB-PESQ} & \textbf{STOI(\%)} & \textbf{WB-PESQ} & \textbf{NB-PESQ} &\textbf{STOI(\%)} \\ \midrule 

Top-$k$ by $KGS$ & \textbf{2.602} & \textbf{3.157} & \textbf{96.10} & \textbf{2.603} & \textbf{3.431} & 93.39 \\
\midrule[1pt]

Top-$k$ by $E^{T}_{p}$ & 2.551 & 3.110 & 95.94 & 2.559 & 3.401 & \textbf{93.46} \\ [2pt] 
Bottom-$k$ by $E^{T}_{p}$ & 2.570 & 3.134 & 96.07 & 2.556 & 3.430 & 93.30 \\ [2pt] 
Top-$k$ by $E^{S}_{p}$ & 2.587 & 3.146 & \textbf{96.10} & 2.577 & 3.418 & 93.38 \\ [2pt] 
Bottom-$k$ by $E^{S}_{p}$ & 2.566 & 3.136 & 95.97 & 2.585 & 3.430 & 93.23 \\  

\bottomrule [1pt]
\end{tabular}
\label{sec:table3}
\vspace{-6pt}
\end{table}
\section{CONCLUSIONS}
\label{sec:typestyle}


This work introduces DISPatch, a KD framework that transfers knowledge only from useful spectrogram patches determined via a KGS, effectively avoiding low‑confidence teacher regions and already reliable areas of the student. DISPatch is simple to plug into existing response-based KD objectives that utilize a spectrogram. Moreover, MSSP respects spectral heterogeneity by using fine-grained low- and coarser high-frequency patches, significantly improving the student's speech enhancement performance. Future work will extend the principle of selective transfer to other speech tasks, such as speech separation and speech restoration, and investigate the suitable top-$k$ setting and a wider range of patch sizes.

\section{Acknowledgements}
This work is supported by the National Research Foundation of Korea (NRF) grant (No. RS-2024-00337945) and STEAM research grant (No. RS-2024-00464269) funded by the Ministry of Science and ICT of Korea government (MSIT), the BK21 FOUR program through the NRF grant funded by the Ministry of Education of Korea government (MOE).

\bibliographystyle{IEEEbib} 
\bibliography{strings,refs}

\end{document}